\documentclass[10pt,conference]{IEEEtran}

\usepackage{amsfonts}
\usepackage[dvips]{graphicx}
\usepackage{times}
\usepackage{cite}
\usepackage{amsmath}
\usepackage{cases}
\usepackage{array}
\usepackage{dsfont}
\usepackage{amssymb}

\usepackage{stfloats}
\usepackage{slashbox}
\usepackage{graphicx}
\usepackage{footnote}
\usepackage{booktabs}
\usepackage{array}
\usepackage{bm}
\usepackage{multicol}
\usepackage{graphicx}

\begin{document}

\title{Generalized Signal Alignment For Arbitrary MIMO Two-Way Relay Channels}
\author{\IEEEauthorblockN{Kangqi Liu$^{*}$, Meixia Tao$^{*}$ and Dingcheng Yang$^{\ddag}$}
\IEEEauthorblockA{$^{*}$Dept. of Electronic Engineering, Shanghai Jiao Tong University, Shanghai, China\\
$^{\ddag}$Information Engineering School, Nanchang University, Nanchang, China\\
Emails: \{forever229272129, mxtao\}@sjtu.edu.cn, yangdingcheng@ncu.edu.cn}
}

\maketitle

\begin{abstract}
In this paper, we consider the arbitrary MIMO two-way relay channels, where there are $K$ source nodes, each equipped with $M_i$ antennas, for $i=1,2,\cdots,K$, and one relay node, equipped with $N$ antennas. Each source node can exchange independent messages with arbitrary other source nodes assisted by the relay. We extend our newly-proposed transmission scheme, \textit{generalized signal alignment} (GSA) in \cite{Liu3}, to arbitrary MIMO two-way relay channels when $N>M_i+M_j$, $\forall i \neq j$. The key idea of GSA is to cancel the interference for each data pair in its specific subspace by two steps. This is realized by jointly designing the precoding matrices at all source nodes and the processing matrix at the relay node. Moreover, the aligned subspaces are orthogonal to each other. By applying the GSA, we show that a necessary condition on the antenna configuration to achieve the DoF upper bound $\min \{\sum_{i=1}^K M_i, 2\sum_{i=2}^K M_i,2N\}$ is $N \geq \max\{\sum_{i=1}^K M_i-M_s-M_t+d_{s,t}\mid \forall s,t\}$. Here, $d_{s,t}$ denotes the DoF of the message exchanged between source node $s$ and $t$. In the special case when the arbitrary MIMO two-way relay channel reduces to the $K$-user MIMO Y channel, we show that our achievable region of DoF upper bound is larger than the previous work.
\end{abstract}

\section{Introduction}
Wireless relay can reduce power, expand coverage and enhance throughput in wireless networks. One of the basic building blocks of relay-aided systems is the two-way relay channel (TWRC), where two source nodes exchange information through a common relay. It promises high spectral efficiency by applying physical layer network coding (PLNC) \cite{Rankov}. Recently, a novel multiple-input multiple-output (MIMO) Y channel was investigated in \cite{Lee1}, where three source nodes exchange independent messages with each other with the help of a common relay. The channel was then extended to $K$-user MIMO Y channel in \cite{Lee}. The authors in \cite{Sezgin} investigated the multi-pair two-way relay channel, where the source nodes are grouped into pairs and the two nodes in each pair exchange independent information with each other. Later, the MIMO two-way X relay channel was investigated in \cite{Xiang}, where all users are divided into two groups and the users in each group exchange independent information with all the users in the other group. More recently, the $K$-user $L$-cluster MIMO multi-way relay channel was proposed in \cite{Tian} to unify the $K$-user MIMO Y channel and the multi-pair two-way relay channel.

A critical metric in characterizing the high signal-to-noise ratio (SNR) performance is degrees of freedom (DoF) \cite{Cadambe}. Interference alignment (IA) has been shown to increase DoF for various wireless multi-user network models \cite{Maddah,Jafar}. IA keeps the interference signals in the smallest number of dimensions, and enables the maximum number of independent data streams to be transmitted. By integrating the concepts of IA and PLNC, signal alignment (SA) for network coding is proposed in \cite{Lee1} for analyzing the DoF of the MIMO Y channel and it is able to achieve
the maximum DoF of the MIMO Y channel when $N>\lceil\frac{3M}{2}\rceil$, where $M$ and $N$ denote the number of antennas at each source node and the relay node, respectively. Then, SA is applied for the analysis of various types of MIMO two-way relay channels \cite{Lee,Xiang,Tian}. However, SA is only applicable for the case when $N<2M$. In our previous work \cite{Liu3}, we proposed a new method, \textit{generalized signal alignment} (GSA), for network coding, which can be applied to align signal pairs even when $N>2M$ for MIMO two-way X relay channel.

In this paper, we propose a more generalized two-way relay channel model, \textit{arbitrary MIMO two-way relay channels}, which unifies various types of MIMO two-way relay channels, where there are $K$ users each equipped with $M_i$, for $i=1,2,\cdots,K$, antennas with $M_1 \geq M_2 \geq \cdots \geq M_K$ and one relay equipped with $N$ antennas. Each user can arbitrarily select one or more partners to conduct independent information exchange. In the special case where each user wants to conduct independent information exchange with all the rest $K-1$ users, the channel model reduces to the $K$-user MIMO Y channel. The MIMO two-way X relay channel is its another special case where each user conducts independent information exchange with the other $\frac{K}{2}$ users. The $K$-user $L$-cluster MIMO multi-way relay channel is also its special case.

Next, we investigate the achievable region of the DoF upper bound for the arbitrary MIMO two-way relay channels with GSA. The main results obtained in this work are as follows:
\begin{itemize}
  \item If $M_i=M$ for $\forall i$, the DoF upper bound of $KM$ is achievable with GSA when $N \geq \frac{(K^2-3K+3)M}{K-1}$, which enlarges the achievable region of the DoF upper bound in \cite{Wang3}.
  \item If $M_1 \geq \sum_{i=2}^K M_i$, the DoF upper bound $2\sum_{i=2}^K M_i$ is achievable with GSA when $N \geq \sum_{i=2}^K M_i$.
  \item If $M_1 \leq \sum_{i=2}^K M_i$, the DoF upper bound $\sum_{i=1}^K M_i$ is achievable with GSA when $N \geq \max\{\sum_{i=1}^K M_i-M_s-M_t+d_{s,t}\mid \forall s,t\}$.
\end{itemize}

Notations: {\bf{Null}} {\bf X} stands for the null space of the matrix ${\bf X}$. $\lfloor x \rfloor$ denotes the largest integer no greater than $x$. $\lceil x \rceil$ denotes the smallest integer no less than $x$. {\bf I} is the identity matrix. ${\bf X}_{[i,j]}$ denotes the element of the matrix ${\bf X}$, which is located at the $i$-th row and $j$-th column. ${\bf X}_{[i,:]}$ denotes the $i$-th row of the matrix. ${\bf X}_{[:,j]}$ denotes the $j$-th column of the matrix.

\section{System Model}
\begin{figure}[t]
\begin{centering}
\rotatebox{90}{\includegraphics[scale=0.6]{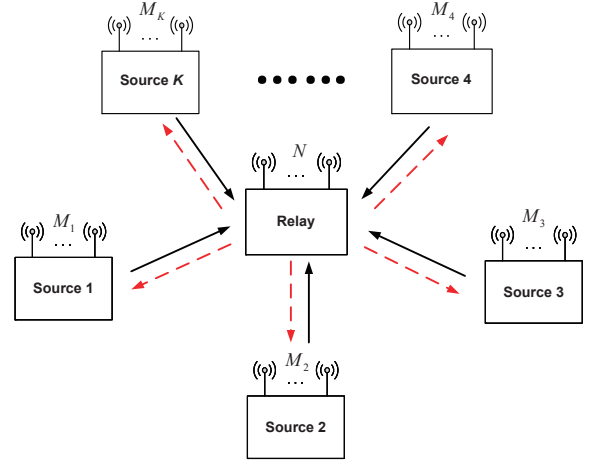}}
\vspace{-0.1cm}
 \caption{Arbitrary MIMO two-way relay channel.}\label{Ge_MIMO_A}
\end{centering}
\vspace{-0.3cm}
\end{figure}

We consider the arbitrary MIMO two-way relay channels as shown in Fig. \ref{Ge_MIMO_A}. It consists of $K$ source nodes, each equipped with $M_i$ antennas, for $i=1,2,\cdots,K$, and one relay node, equipped with $N$ antennas. Each source node can exchange independent messages with arbitrary other source nodes assisted by the relay. Note that source nodes and users are interchangeable in this paper. The independent message transmitted from source node $i$ to source node $j$, if any, is denoted as $W_{i,j}$. At each time slot, the message is encoded into a $d_{i,j} \times 1$ symbol vector ${\textbf{s}}_{i,j}=[s_{i,j}^1,s_{i,j}^2,\cdots,s_{i,j}^{d_{i,j}}]^T$, where $d_{i,j}$ denotes the number of independent data streams transmitted from source $i$ to source source $j$. If there is no information exchange between source node $i$ and source node $j$, we have $d_{i,j} = 0$. We define a $K \times K$ matrix ${\bf D}$ named as \textit{data switch matrix} as
\begin{equation}\label{D}
{{\bf D}}=\left[\begin{array}{ccccc}
                0 & d_{1,2} & d_{1,3} & \cdots & d_{1,K} \\
                d_{2,1} & 0 & d_{2,3} & \cdots & d_{2,K} \\
                d_{3,1} & d_{3,2} & 0 & \cdots & d_{3,K} \\
                \vdots & \vdots & \vdots & \ddots & \vdots \\
                d_{K,1} & d_{K,2} & d_{K,3} & \cdots & 0
              \end{array}
\right],
\end{equation}
where ${\bf D}_{[i,j]}$ denotes $d_{i,j}$. Clearly, the diagonal elements of the matrix ${\bf D}$ are zero.

Let ${\cal K}_i$ denote the set of users that source node $i$ will exchange information with. Taking source node $1$ for example, the transmitted signal vector ${{\bf x}}_1$ from source node $1$ is given by
\begin{eqnarray}\label{x_1}
{\bf x}_1 =\sum\limits_{j \in {\cal K}_1} {\bf V}_{1,j}{\textbf{s}}_{1,j}= {{\bf V}}_1{\textbf{s}}_1,
\end{eqnarray}
where ${\bf V}_{1,j}$ is the $M_1 \times d_{1,j}$ precoding matrix for the information symbol vector ${\bf s}_{1,j}$ to be sent to source node $j$, ${\bf V}_1$ is row vector consisting of all ${\bf V}_{1,j}$ and $\textbf{s}_1$ is the column vector consisting of all $\textbf{s}_{1.j}$, for $j \in {\cal K}_1$.

The communication of the total messages takes place in two phases: the multiple access (MAC) phase and the broadcast (BC) phase. In the MAC phase, all $K$ source nodes transmit their signals to the relay simultaneously. The received signal ${\bf y}_r$ at the relay is given by
\begin{eqnarray}\label{y_r}
{{\bf y}}_r=\sum\limits_{i=1}^{K}{{\bf H}}_{i,r}{{\bf x}}_{i}+{{\bf n}}_r
\end{eqnarray}
where ${{\bf H}}_{i,r}$ denotes the frequency-flat quasi-static $N \times M_i$ complex channel matrix from source node $i$ to the relay and ${{\bf n}}_r$ denotes the $N\times 1$ additive white Gaussian noise (AWGN) with variance $\sigma_n^2$.

Upon receiving ${{\bf y}}_r$ in \eqref{y_r}, the relay processes it to obtain a mixed signal ${\bf x}_r$,  and broadcasts to all the users. The received signal at source node $i$ can be written as
\begin{eqnarray}\label{y_i}
{{\bf y}}_i={\bf G}_{r,i}{\bf x}_r+{{\bf n}}_i
\end{eqnarray}
where ${{\bf G}}_{r,i}$ denotes the frequency-flat quasi-static $M_i \times N$ complex channel matrix from the relay to the source node $i$, and ${\bf n}_i$ denotes the AWGN at the node $i$. Each user tries to obtain its desirable signal from its received signal using its own transmit signal as side information.

To pursue the performance limits, we assume that the channel state information $\{{\bf H}_{i,r}\}$ and $\{{\bf G}_{r,i}\}$ are perfectly known at all source nodes and relay, following the convention in \cite{Lee1,Lee,Wang3,Xiang,Tian}. The entries of the channel matrices and those of the noise vector ${{\bf n}}_r$, ${\bf n}_i$ are independent and identically distributed (i.i.d.) zero-mean complex Gaussian random variables with unit variance. Thus, each channel matrix is full rank with probability $1$.

\section{Generalized Signal Alignment}
In this section, we review the principle of \textit{generalized signal alignment} proposed in our previous work \cite{Liu3} and present it in a more general manner.

Without loss of generality, we assume that $M_1 \geq M_2 \geq \cdots \geq M_K$. It is worth mentioning that the total DoF upper bound of the arbitrary MIMO two-way relay channel is $\min \{\sum_{i=1}^K M_i, 2\sum_{i=2}^K M_i,2N\}$ by applying cut-set theorem \cite{Cover} and genie-aided method \cite{Chaaban}. In this paper, we are interested in the case when $N>M_i+M_j$, for $\forall i,j$, where SA is not applicable and analyze the achievability of the upper bound.

The DoF upper bound of the source node $i$ is
\begin{align}\label{d_per}\nonumber
d_i&=\\\nonumber
\sum\limits_{k=1,k \neq i}^K d_{i,k} &\leq \min\{\underbrace{\min\{M_i,N\}}_{source~node~i~to~relay},\underbrace{\min\{\sum\limits_{j=1,j \neq i}^K M_j,N\}}_{relay~to~others}\}\\
&=\min\{M_i,\sum\limits_{j=1,j \neq i}^K M_j,N\}.
\end{align}
Due to the constraints of $N>M_i+M_j$ and $M_1 \geq M_2 \geq \cdots \geq M_K$, Eq. \eqref{d_per} reduces to $d_1 \leq \min \{M_1, \sum_{j=2}^K M_j\}$ and $d_i \leq M_i$ for source node $i$, $\forall i \neq 1$. The total DoF upper bound of the arbitrary MIMO two-way relay channel reduces to $\min \{\sum_{i=1}^K M_i, 2\sum_{i=2}^K M_i\}$ for this case. As for convenience, we define $\tilde{M}_1$ as $\min \{M_1, \sum_{j=2}^K M_j\}$. We only utilize $\tilde{M}_1$ antennas at the source node $1$, where antenna deactivation \cite{Wang3} is applied for analysis.

To achieve this DoF upper bound, we have two assumptions.
\begin{itemize}
    \item The number of data streams from source node $i$ transmitted to source node $j$ is equal to the number of data streams from source node $j$ transmitted to source node $i$. This is equivalent to that the data switch matrix {\bf D} is symmetric.
    \item The DoF of the source node $i$ is $M_i$ (when $i=1$, it is $\tilde{M}_1$). This is equivalent to that the sum of each row of the matrix {\bf D} is $M_i$, yielding a total DoF of $\tilde{M}_1+\sum_{j=2}^K M_j$.
\end{itemize}

In the MAC phase, each source node transmits the precoded signals to the relay simultaneously. We rewrite the received signal \eqref{y_r} as
\begin{align}\nonumber
&{{\bf y}}_r\\\nonumber
&=\left[{\bf H}_{1,r}~{\bf H}_{2,r}~\cdots~{\bf H}_{K,r}\right]\left[\begin{array}{cccc}
                                                                                       {\bf V}_1 & \textbf{0} & \textbf{0} & \textbf{0} \\
                                                                                       \textbf{0} & {\bf V}_2 & \textbf{0} & \textbf{0} \\
                                                                                       \textbf{0} & \textbf{0} & \ddots & \vdots \\
                                                                                       \textbf{0} & \textbf{0} & \cdots & {\bf V}_K
                                                                                     \end{array}
\right]\left[\begin{array}{c}\textbf{s}_1\\\textbf{s}_2\\\vdots\\\textbf{s}_K\end{array}\right]\\\nonumber
&~~~+{\bf n}_r\\\label{y_r_generalized}
&={\bf H}{\bf V}\textbf{s}+{\bf n}_r,
\end{align}
where ${\bf H}$ is the overall channel matrix, ${\bf V}$ is the block-diagonal overall precoding matrix and $\textbf{s}$ is the transmitted signal vector for all the source nodes.

When $N \geq \tilde{M}_1+\sum_{i=2}^K M_i$, the relay can decode all the $\tilde{M}_1+\sum_{i=2}^K M_i$ data streams and the decode-and-forward (DF) relay is the optimal transmission strategy. When $N < \tilde{M}_1+\sum_{i=2}^K M_i$, it is impossible for the relay to decode all the $\tilde{M}_1+\sum_{i=2}^K M_i$ data streams. However, considering the idea of physical layer network coding, we only need to obtain the following network-coded symbol vector at the relay
\begin{eqnarray}\label{s_oplus_generalized}
\textbf{s}_{\oplus}=\left[\textbf{s}_{1,2}+\textbf{s}_{2,1}, \cdots, \textbf{s}_{i,j}+\textbf{s}_{j,i}, \cdots, \textbf{s}_{K-1,K}+\textbf{s}_{K,K-1}\right]^T.
\end{eqnarray}
Note that the terms $\textbf{s}_{i,j}+\textbf{s}_{j,i}$ exists if and only if $d_{i,j} \neq 0$.

Since $N>M_i+M_j$, for $\forall i,j$, it is clear that $\textbf{s}_{\oplus}$ cannot be obtained directly by designing ${\bf V}$ in this case only due to the SA condition. Joint design of the source precoding matrix and relay projection matrix should be considered. Let $\bf A$ denote the projection matrix at the relay. Then, mathematically, the signal after projection is
\begin{align}\label{y_r_projection}
\hat{{\bf y}}_r={{\bf A}}{{\bf y}}_r={{\bf A}}{\bf H}{\bf V}\textbf{s}+{{\bf A}}{\bf n}_r=\textbf{s}_{\oplus}+{{\bf A}}{\bf n}_r.
\end{align}

Here, some notations and terms need to be introduced first. Let $(s_{i,j}^d, s_{j,i}^d)$ denote the $d$-th pair of data streams exchanged between the source pair ($i$, $j$), for $d=1,~\cdots,~d_{i,j}$. For each $(s_{i,j}^d, s_{j,i}^d)$, we treat all the other signals as interference and divide them into two categories, external interference and internal interference. The external interference is composed by the signals transmitted from all other sources except source $i$ and source $j$, namely {$s_{m,n}^l, \forall m \neq i,j$}. The internal interference comes from all the other signals transmitted by source node $i$ and source node $j$, namely $\{s_{i,n}^l \mid n \neq j \} \cup \{s_{j,n}^l \mid n \neq i \} \cup \{s_{i,j}^l \mid l \neq d \} \cup \{s_{j,i}^l \mid l \neq d \}$.

The target is to align each pair of signals to be exchanged in a same subspace. For this purpose, we project all other signals in the null of the target subspace of each signal pair. Then, for each subspace, there only remains the target signal pair without any interference. Effectively, every signal pair is naturally aligned in its respective subspace. From this perspective, we name this signal processing method as \textit{generalized signal alignment}. The key idea of GSA involves two steps:
\begin{itemize}
  \item Design the projection matrix $\bf A$ at the relay so as to cancel the external interference for each data pair in its specific subspace.
  \item Design the precoding matrix $\bf V$ at each source node so as to cancel the internal interference for each data pair in its specific subspace.
\end{itemize}

Note that $\bf A$ does not always exist, we will analyze the necessary condition in the next section.

During the BC phase, the relay broadcasts an estimate of $\textbf{s}_{\oplus}$ using the precoding matrix ${{\bf U}}$. We rewrite the received signal \eqref{y_i} as
\begin{eqnarray}\label{y_i_generalized}
{{\bf y}}_i={\bf G}_{r,i}{\bf U}\textbf{s}_{\oplus}+{{\bf n}}_i
\end{eqnarray}

Then each source node can obtain its desired message from the received message using its own transmit signal as the side information.

\section{DoF Achievability with Generalized Signal Alignment}
In this section, we investigate the DoF achievability with \textit{generalized signal alignment}.

\subsection{$M_i=M$,~for~$\forall i$}
\textit{Theorem 1}: If $M_i=M$, for $\forall i$, and $N \geq \frac{(K^2-3K+3)M}{K-1}$, there exists a $KM \times KM$ block-diagonal precoding matrix {\bf V} and a $\frac{KM}{2} \times N$ projection matrix {\bf A} such that
\begin{align}\label{AVdesign}
{{\bf A}}{\bf H}{\bf V}\textbf{s}=\textbf{s}_{\oplus},
\end{align}
and the DoF of $KM$ is achievable for the arbitrary MIMO two-way relay channels with $d_{i,j} \leq N-(K-2)M$, for $\forall i,j$.
\begin{proof}
From \textit{assumption 2} in the above section, we have
\begin{align}\label{AVdesign}
d_i=\sum\limits_{j \in {\cal K}_i} d_{i,j}=M_i=M.
\end{align}
It is obvious that $\max\{d_{i,j}\} \geq \frac{M}{K-1}$, for $\forall i,j$. Here, the equality holds if and only if $d_{i,j} = \frac{M}{K-1}$, for $\forall i,j$, which is the special case of the arbitrary MIMO two-way relay channels, $K$-user MIMO Y channel. In other words, $K$-user MIMO Y channel requires the minimum number of antennas at relay to achieve the DoF of $KM$ compared to other arbitrary MIMO two-way relay channels in the same condition at the source nodes. We first analyze the DoF achievability of the $K$-user MIMO Y channel with GSA. The data switch matrix ${\bf D}$ of the $K$-user MIMO Y channel is
\begin{equation}\label{D_Y}
{{\bf D}}=\left[\begin{array}{ccccc}
                 0 & \frac{M}{K-1} & \cdots & \frac{M}{K-1} & \frac{M}{K-1} \\
                 \frac{M}{K-1} & 0 & \cdots & \frac{M}{K-1} & \frac{M}{K-1} \\
                 \vdots & \vdots & \ddots & \vdots & \vdots \\
                 \frac{M}{K-1} & \frac{M}{K-1} & \cdots & 0 & \frac{M}{K-1} \\
                 \frac{M}{K-1} & \frac{M}{K-1} & \cdots & \frac{M}{K-1} & 0
               \end{array}
\right].
\end{equation}

\textbf{Case 1: $M$ is divisible by $K-1$.} ~~In this case, $d_{i,j}$ is exactly $\frac{M}{K-1}$ for $\forall i \neq j$. We achieve the DoF of $KM$ in three steps.

\textbf{Step 1: Design the projection matrix so as to cancel the external interference of each pair $(s_{i,j}^d, s_{j,i}^d)$ in its specific subspace.} Denoting ${\bf A}_p$ as the $\frac{(i-1)M}{K-1}+1$ to the $\frac{iM}{K-1}$ row vectors of {\bf A}, for $p=1, 2, \cdots, \frac{K(K-1)}{2}$. Each ${\bf A}_p$ can be thought as a projection matrix for the transmitted signals of a source node pair, totally $\frac{K(K-1)}{2}$ pairs. Thus, we design ${\bf A}_p$ as follows to cancel the external interference of each pair $(s_{i,j}^d, s_{j,i}^d)$.
\begin{eqnarray}\nonumber\label{a_1234}
{\bf A}_1^T&\subseteq \textbf{Null}~\big[{\bf H}_{3,r}~{\bf H}_{4,r}~\cdots~{\bf H}_{K,r}\big]^T~\\\nonumber
{\bf A}_2^T&\subseteq \textbf{Null}~\big[{\bf H}_{2,r}~{\bf H}_{4,r}~\cdots~{\bf H}_{K,r}\big]^T~\\\nonumber
\vdots\\\nonumber
\vdots\\
{\bf A}_{\frac{K(K-1)}{2}}^T&\subseteq \textbf{Null}~\big[{\bf H}_{1,r}~{\bf H}_{2,r}~\cdots~{\bf H}_{K-2,r}\big]^T.
\end{eqnarray}

Here, ${\bf A}_1$ is for source pair (1,2), ${\bf A}_2$ is for source pair (1,3) and ${\bf A}_{\frac{K(K-1)}{2}}$ is for source pair ($K-1$, $K$) and each row vector ${\bf A}_{p_{[l,:]}}$ is for source pair $(s_{i,j}^l, s_{j,i}^l)$. Each ${\bf A}_p$ contains $\frac{M}{K-1}$ rows which align $\frac{M}{K-1}$ streams to $\frac{M}{K-1}$ orthogonal subspace.

We can see that ${\bf A}_p^{T}$ for pair $(i,j)$ is an $N \times \frac{M}{K-1}$ matrix and it locates in the null space of the corresponding matrix as $\big[{\bf H}_{1,r}~\cdots~{\bf H}_{i-1,r}~{\bf H}_{i+1,r}~\cdots~{\bf H}_{j-1,r}~{\bf H}_{j+1,r}~\cdots~{\bf H}_{K,r}\big]^T$, which is a $(K-2)M \times N$ matrix. The matrix ${\bf A}_p^{T}$ exists if and only if $N-(K-2)M \geq d_{i,j}=\frac{M}{K-1}$, which is equivalent to $N \geq \frac{(K^2-3K+3)M}{K-1}$.

\textbf{Step 2: Design the precoding matrix at each source node so as to cancel the internal interference of each pair $(s_{i,j}^d, s_{j,i}^d)$ in its specific subspace.} After obtaining the matrix {\bf A}, we need to design the precoding matrix to rotate the vectors of each signal pairs to orthogonal. Let ${\bf C}={\bf A}{\bf H}$. We denotes the nonzero rows of the first $M$ column as ${\bf C}_1$, and similar notions for ${\bf C}_i$, $i=1,2,\cdots,K$. We can design the matrix ${\bf V}_i={\bf C}_i^{-1}$. It is worth mentioning that each ${\bf C}_i$ is a full-rank $M \times M$ square matrix because the independence of each ${\bf A}_{p_{[i,:]}}^T$. Hence, ${\bf V}_i={\bf C}_i^{-1}$ always exists. After designing ${\bf V}_i$, the internal interference has been cancelled in the specific subspace, which yields the alignment of each two signals.

\textbf{Step 3: Design the precoding matrix at relay for broadcasting.} We use the method of interference nulling to design the precoding matrix {\bf U}. We can write {\bf U} as follows.
\begin{equation}\label{UU}
{\bf U}=
\left[
\begin{array}{ccc}
 {\bf U}_{1}~{\bf U}_{2}~\cdots~{\bf U}_{\frac{K(K-1)}{2}}
\end{array}
\right],
\end{equation}
where each ${\bf U}_p$ is an $N \times \frac{M}{K-1}$ matrix and
\begin{eqnarray}\nonumber\label{UU3}
{\bf U}_1&\subseteq \textbf{Null}~\big[{\bf G}_{r,3}^T~{\bf G}_{r,4}^T~\cdots~{\bf G}_{r,K}^T\big]^T~\\\nonumber
{\bf U}_2&\subseteq \textbf{Null}~\big[{\bf G}_{r,2}^T~{\bf G}_{r,4}^T~\cdots~{\bf G}_{r,K}^T\big]^T~\\\nonumber
\vdots\\\nonumber
\vdots\\
{\bf U}_{\frac{K(K-1)}{2}}&\subseteq \textbf{Null}~\big[{\bf G}_{r,1}^T~{\bf G}_{r,2}^T~\cdots~{\bf G}_{r,K-2}^T\big]^T
\end{eqnarray}

$\big[{\bf G}_{r,1}^T~\cdots~{\bf G}_{r,i-1}^T~{\bf G}_{r,i+1}^T~\cdots~{\bf G}_{r,j-1}^T~{\bf G}_{r,j+1}^T~\cdots~{\bf G}_{r,K}^T\big]^T$ is a $(K-2)M \times N$ matrix. Then the matrix ${\bf U}_p$ exists if and only if $N-(K-2)M \geq \frac{M}{K-1}$, which is equivalent to $N \geq \frac{(K^2-3K+3)M}{K-1}$. Hence, we can apply GSA-based transmission scheme when $M$ is divisible by $K-1$ and $N \geq \frac{(K^2-3K+3)M}{K-1}$ to achieve the DoF upper bound $KM$ for $K$-user MIMO Y channel.

\textbf{Case 2: $M$ is not divisible by $K-1$.} In this case, the number of data streams exchanged with each pair is $\frac{M}{K-1}$, which is a fraction. We use the idea of the symbol extension \cite{Jafar1} to prove the achievability of this DoF upper bound. In the previous research, using symbol extension, the achievable DoF of the two user MIMO X channel is enlarged from the $\lfloor\frac{4M}{3}\rfloor$ DoF \cite{Maddah} to $\frac{4M}{3}$ \cite{Jafar}. Here, we introduce how to achieve the DoF of $KM$ using both symbol extension and GSA.

We consider $(K-1)$-symbol extensions of the channel model, where the channel coefficients varying over time are unnecessary here. The received signal at the relay can be written as
\begin{align}\nonumber
{\bf y}_r&=\left[\begin{array}{c}{\bf y}_r(1)\\{\bf y}_r(2)\\\vdots\\{\bf y}_r(K-1)\end{array}\right]\\\nonumber
&=\left[\begin{array}{cccc}
{\bf H}(1) & {\bf 0} & \cdots & {\bf 0} \\
{\bf 0} & {\bf H}(2) & \cdots & {\bf 0} \\
\vdots & \vdots & \ddots & {\bf 0} \\
{\bf 0} & {\bf 0} & \cdots & {\bf H}(K-1)
\end{array}
\right]\left[\begin{array}{c}{\bf x}(1)\\{\bf x}(2)\\\vdots\\{\bf x}(K-1)\end{array}\right]\\\nonumber
&+\left[\begin{array}{c}{\bf n}_r(1)\\{\bf n}_r(2)\\\vdots\\{\bf n}_r(K-1)\end{array}\right]\\
&={\bf H}^{\S}{\bf x}^{\S}+{\bf n}_r^{\S}.
\end{align}
where ${\bf y}_r(t)$, ${\bf H}(t)$, ${\bf x}(t)$ and ${\bf n}_r(t)$ denote the $t$-th time/frequency slot of received signal, channel matrices, transmitted signals and noise, ${\bf H}^{\S}$ denotes the equivalent channel matrix, ${\bf x}^{\S}$ denotes the equivalent transmitted signals and ${\bf n}_r^{\S}$ denotes the equivalent noise.

Note that ${\bf H}^{\S}$ is a $(K-1)N\times(K-1)KM$ matrix. The system model is equivalent to the $K$-user MIMO Y channel, where each source node is equipped with $(K-1)M$ antennas and the relay is equipped with $(K-1)N$ antennas. The equivalent number of the source node is a multiple of $K-1$. It turns to be \textbf{Case 1} and we can then apply GSA to achieve the DoF $(K-1)KM$ over $(K-1)$-symbol extensions. This fact implies that the DoF of $KM$ is achievable in the original K-user MIMO Y channel. The antenna constraints can be written as
\begin{equation}\label{s_NKMYYY}
(K-1)N-(K-2)(K-1)M \geq \frac{(K-1)M}{K-1}
\end{equation}

That is
\begin{equation}\label{s_NKMYYYY}
N \geq \frac{(K^2-3K+3)M}{K-1}
\end{equation}

The above analysis shows that the generalized signal alignment based transmission scheme can achieve the DoF of $KM$ with $N \geq \frac{(K^2-3K+3)M}{K-1}$ when $N>2M$ in $K$-user MIMO Y channel. On the other hand, it is clear to see that under the condition $N \geq \frac{(K^2-3K+3)M}{K-1} \geq \frac{KM}{2}$. $KM$ is also the the DoF upper bound of the $K$-user MIMO Y channel.

If $\max\{d_{i,j}\} > \frac{M}{K-1}$, for $\forall i,j$, the channel is no longer the $K$-user MIMO Y channel. Due to the space limitation, we will give the rest of the proof together with \textit{Theorem 3} as a special case.
\end{proof}

\begin{figure}[t]
\begin{centering}
\includegraphics[scale=0.45]{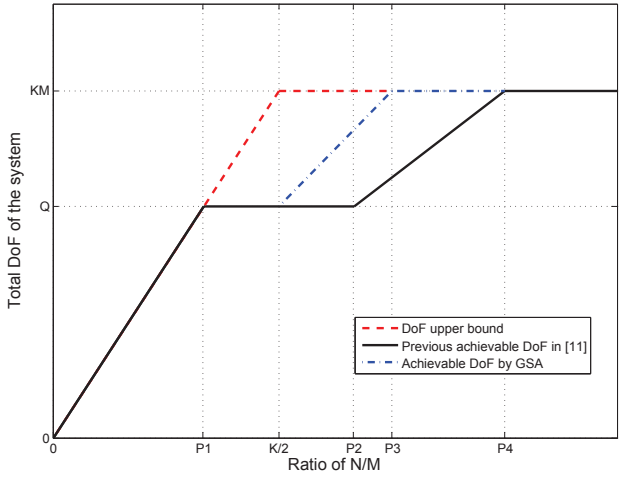}
\vspace{-0.1cm}
 \caption{Achievable DoF upper bound region for $K$-user MIMO Y channel. ($P1=\frac{2K^2-2K}{K^2-K+2}$, $P2=\frac{4K^2-8K}{K^2-K+2}$, $P3=\frac{K^2-3K+3}{K-1}$, $P4=\frac{K^2-2K}{K-1}$ and $Q=\frac{(4K^2-4K)M}{K^2-K+2}$)}\label{DoF_GeY}
\end{centering}
\vspace{-0.3cm}
\end{figure}

Note that the authors in \cite{Wang3} showed that the upper bound $KM$ of DoF is achievable in the case when $N \geq \frac{K^2-2K}{K-1}$, which is a subset of our region. Fig. \ref{DoF_GeY} illustrates the achievable DoF for different antenna configurations.

\subsection{$M_1 \geq \sum_{i=2}^K M_i$}
\textit{Theorem 2}: If $M_1 \geq \sum_{i=2}^K M_i$, i.e. $\tilde{M}_1 = \sum_{i=2}^K M_i$, the DoF upper bound $2\sum_{i=2}^K M_i$ is achievable when $N \geq \sum_{i=2}^K M_i$.
\begin{proof}
To achieve this upper bound, the data switch matrix ${\bf D}$ is unique here, which can be written as
\begin{equation}\label{D_Y2}
{{\bf D}}=\left[\begin{array}{ccccc}
                 0 & M_2 & \cdots & M_{K-1} & M_K \\
                 M_2 & 0 & \cdots & 0 & 0 \\
                 \vdots & \vdots & \ddots & \vdots & \vdots \\
                 M_{K-1} & 0 & \cdots & 0 & 0 \\
                 M_K & 0 & \cdots & 0 & 0
               \end{array}
\right].
\end{equation}
If $d_{i,j} > 0$, for $\forall i,j \neq 1$, then $d_{1,j} \leq M_j-d_{i,j}< M_j$. This leads to $d_1<\sum_{i=2}^K M_i$, where the DoF upper bound $2\sum_{i=2}^K M_i$ is not achievable. Hence, the DoF upper bound $2\sum_{i=2}^K M_i$ is achievable only under the data switch matrix shown in Eq. \eqref{D_Y2}.

We use both antenna deactivation and GSA this time, where source node 1 only utilizes $\sum_{i=2}^K M_i$ antennas. Compared to \textit{Theorem 1}, the key idea of designing $\bf A$, $\bf V$ and $\bf U$ is similar. $\bf A$ and $\bf V$ are designed to cancel the external and internal interference of each pair $(s_{i,j}^d, s_{j,i}^d)$ in its specific subspace. $\bf U$ is designed for broadcasting. The only difference is that ${\bf A}_i$ is an $M_{i+1} \times N$ submatrix of {\bf A}, i.e. ${\bf A}_i$ is for pair $(1,~i+1)$. In specific, the matrix ${\bf A}_i^{T}$ exists if and only if $N-\sum_{j=2}^{i-1} M_j-\sum_{j=i+1}^{K} M_j\geq M_i$, which is equivalent to $N \geq \sum_{i=2}^K M_i$. ${\bf U}_i$ also exists if and only if $N-\sum_{j=2}^{i-1} M_j-\sum_{j=i+1}^{K} M_j\geq M_i$, which is equivalent to $N \geq \sum_{i=2}^K M_i$. Due to space limitations, we omit the detailed proof here.
\end{proof}

\subsection{$M_1 \leq \sum_{i=2}^K M_i$}
\textit{Theorem 3}: If $M_1 \leq \sum_{i=2}^K M_i$, i.e. $\tilde{M}_1 = M_1$, the DoF upper bound $\sum_{i=1}^K M_i$ is achievable when $N \geq \max\{\sum_{i=1}^K M_i-M_s-M_t+d_{s,t}\mid \forall s,t\}$.
\begin{proof}
The key idea to achieve the DoF of $\sum_{i=1}^K M_i$ is also to design $\bf A$, $\bf V$ and $\bf U$. We cannot give a close-form solution for the minimum of $N$ this time. As for clarity, we give an example for the antenna constraints for pair ($s$,$t$) to achieve the DoF of $d_{s,t}$. The projection matrix ${\bf A}_i$ for pair ($s$,$t$) should satisfy
\begin{equation}\label{a_i}
{\bf A}_i^T\subseteq \textbf{Null}~\big[{\bf H}_{1,r}~\cdots~{\bf H}_{s-1,r}~{\bf H}_{s+1,r}~\cdots~{\bf H}_{t-1,r}~{\bf H}_{t+1,r}~{\bf H}_{K,r}\big]^T.
\end{equation}
Note that ${\bf A}_i^T$ is an $N \times d_{s,t}$ matrix and $\big[{\bf H}_{1,r}~\cdots~{\bf H}_{s-1,r}~{\bf H}_{s+1,r}~\cdots~{\bf H}_{t-1,r}~{\bf H}_{t+1,r}~{\bf H}_{K,r}\big]^T$ is a $(\sum_{i=1}^K M_i-M_s-M_t) \times N$ matrix. Hence, ${\bf A}_i^T$ for pair ($s$,$t$) exists if and only if $N \geq \sum_{i=1}^K M_i-M_s-M_t+d_{s,t}$. After designing ${\bf A}$, the external interference has been cancelled in the specific subspace.

Let ${\bf C}={\bf A}{\bf H}$. We denotes the nonzero rows of the first $M_1$ column as ${\bf C}_1$, and similar notions for ${\bf C}_i$. We can design the matrix ${\bf V}_i={\bf C}_i^{-1}$. It is worth mentioning that each ${\bf C}_i$ is a full-rank $M_i \times M_i$ square matrix because of the independence of each ${\bf A}_{[i,:]}$. Hence, ${\bf C}_i^{-1}$ always exists. After designing ${\bf V}$, the internal interference has been cancelled in the specific subspace, which yields the alignment of each two signals.

The broadcast precoding matrix ${\bf U}_i$ for pair ($s$,$t$) should satisfy
\begin{equation}\label{a_i}
{\bf U}_i\subseteq \textbf{Null}~\big[{\bf G}_{r,1}^T~\cdots~{\bf G}_{r,s-1}^T~{\bf G}_{r,s+1}^T~\cdots~{\bf G}_{r,t-1}^T~{\bf G}_{r,t+1}^T~{\bf G}_{r,K}^T\big]^T.
\end{equation}
Note that ${\bf U}_i$ is an $N \times d_{s,t}$ matrix and $\big[{\bf G}_{r,1}^T~\cdots~{\bf G}_{r,s-1}^T~{\bf G}_{r,s+1}^T~\cdots~{\bf G}_{r,t-1}^T~{\bf G}_{r,t+1}^T~{\bf G}_{r,K}^T\big]^T$ is an $(\sum_{i=1}^K M_i-M_s-M_t) \times N$ matrix. Hence, ${\bf U}_i$ for pair ($s$,$t$) exists if and only if $N \geq \sum_{i=1}^K M_i-M_s-M_t+d_{s,t}$.

This condition is required for every signal pair. Hence, the projection matrix $\bf A$ and the broadcast precoding matrix $\bf U$ exist if and only if $N \geq \max\{\sum_{i=1}^K M_i-M_s-M_t+d_{s,t}\mid \forall s,t\}$.
\end{proof}

For the rest proof of \textit{Theorem 1}, we notice that it is the special case of the \textit{Theorem 3}. Due to the assumption of $M_i=M$, for $\forall i$, we can simplify the condition $N \geq \max\{\sum_{i=1}^K M_i-M_s-M_t+d_{s,t}\mid \forall s,t\}$ to $N \geq (K-2)M+\max\{d_{s,t}\mid \forall s,t\} \geq (K-2)M+\frac{M}{K-1}=\frac{(K^2-3K+3)M}{K-1}$. The achievable DoF is $\sum_{i=1}^K M_i=KM$. Hence, \textit{Theorem 1} has been proved completely.

\subsection{Minimum antenna number required at the relay for specific data switch matrix}
In this subsection, we investigate the minimum number of antennas required at the relay for specific data switch matrix. Given the $K \times K$ data switch matrix ${\bf D}$ as \eqref{D}. Based on \textit{Theorems 1, 2, 3}, we summarize the algorithm for finding the minimum number of antennas required at relay according to \eqref{D} in the following chart.

\vspace{0.4cm} \hrule \hrule \vspace{0.1cm}
\begin{center}
\textbf{Algorithm}
\end{center}
 \vspace{0.1cm} \hrule
\vspace{0.2cm} ~~
1: \textbf{for} $i=1$ to $K$
\vspace{0.1cm} ~~

2: ~\textbf{for} $j=1$ to $K$
\vspace{0.1cm} ~~

3: ~~\textbf{if} $d_{i,j} \neq 0$
\vspace{0.1cm} ~~

4: ~~~Set {\bf T}={\bf D}
\vspace{0.1cm} ~~

5: ~~~Delete the $i$-th row and the $j$-th column of {\bf T}
\vspace{0.1cm} ~~

6: ~~~Sum the elements of {\bf T} and denote it as $N_{i,j}$
\vspace{0.1cm} ~~

7: ~~\textbf{end if}
\vspace{0.1cm} ~~

8: ~\textbf{end for} $(j)$
\vspace{0.1cm} ~~

9: \textbf{end for} $(i)$
\vspace{0.1cm} ~~

10: Find $N=\max\{N_{i,j}\}$, for $\forall i,j$

\vspace{0.2cm} \hrule \vspace{0.4cm}

Note that \textbf{Step 4-6} is based on the constraints of $N \geq \sum_{i=1}^K M_i-M_s-M_t+d_{s,t}$. Then $N$ is the minimum antennas required at relay.

\section{Conclusion}
In this paper, we have analyzed the achievability of the DoF upper bound for arbitrary MIMO two-way relay channels when $N>M_i+M_j$ for $\forall i \neq j$. In the newly-proposed GSA transmission scheme, the projection matrix at the relay and the precoding matrix at the source nodes are designed jointly so that the signals to be exchanged between each source node pair are aligned at the relay. The whole process of the alignment is separated into two steps, external interference cancellation and internal interference cancellation. We show that $N \geq \max\{\sum_{i=1}^K M_i-M_s-M_t+d_{s,t}\mid \forall s,t\}$ is the necessary condition to achieve the upper bound of the total DoF of $\min \{\sum_{i=1}^K M_i, 2\sum_{i=2}^K M_i,2N\}$.

\bibliographystyle{IEEEtran}
\bibliography{IEEEabrv,reference}

\begin{thebibliography}{10}
\providecommand{\url}[1]{#1}
\csname url@samestyle\endcsname
\providecommand{\newblock}{\relax}
\providecommand{\bibinfo}[2]{#2}
\providecommand{\BIBentrySTDinterwordspacing}{\spaceskip=0pt\relax}
\providecommand{\BIBentryALTinterwordstretchfactor}{4}
\providecommand{\BIBentryALTinterwordspacing}{\spaceskip=\fontdimen2\font plus
\BIBentryALTinterwordstretchfactor\fontdimen3\font minus
  \fontdimen4\font\relax}
\providecommand{\BIBforeignlanguage}[2]{{%
\expandafter\ifx\csname l@#1\endcsname\relax
\typeout{** WARNING: IEEEtran.bst: No hyphenation pattern has been}%
\typeout{** loaded for the language `#1'. Using the pattern for}%
\typeout{** the default language instead.}%
\else
\language=\csname l@#1\endcsname
\fi
#2}}
\providecommand{\BIBdecl}{\relax}
\BIBdecl

\bibitem{Liu3}
K.~Liu, M.~Tao, Z.~Xiang, and X.~Long, ``Generalized signal alignment for
  {MIMO} two-way {X} relay channels,'' in \emph{to appear in IEEE International
  Conference on Communications (ICC), http://arxiv.org/abs/1402.1607}, 2014.

\bibitem{Rankov}
B.~Rankov and A.Wittneben, ``Spectral efficient protocols for half-duplex
  fading relay channels,'' \emph{IEEE Journal on Selected Areas in
  Communications}, vol.~25, no.~2, Feb. 2007.

\bibitem{Lee1}
N.~Lee, J.~Lee, and J.~Chun, ``Degrees of freedom on the {MIMO Y} channel:
  signal space alignment for network coding,'' \emph{IEEE Transactions on
  Information Theory}, vol.~56, no.~7, pp. 3332 -- 3342, Jul. 2010.

\bibitem{Lee}
K.~Lee, N.~Lee, and I.~Lee, ``Achievable degrees of freedom on {K-user} {Y}
  channels,'' \emph{IEEE Transactions on Wireless Communications}, vol.~11,
  no.~3, pp. 1210 -- 1219, Mar. 2012.

\bibitem{Sezgin}
A.~Sezgin, A.~S. Avestimehr, M.~A. Khajehnejad, and B.~Hassibi,
  ``Divide-and-conquer: Approaching the capacity of the two-pair bidirectional
  gaussian relay network,'' \emph{IEEE Transactions on Information Theory},
  vol.~58, no.~4, pp. 2434--2454, Apr. 2012.

\bibitem{Xiang}
Z.~Xiang, M.~Tao, J.~Mo, and X.~Wang, ``Degrees of freedom for {MIMO} two-way
  {X} relay channel,'' \emph{IEEE Transactions on Signal Processing}, vol.~61,
  pp. 1711 -- 1720, 2013.

\bibitem{Tian}
Y.~Tian and A.~Yener, ``Degrees of freedom for the {MIMO} multi-way relay
  channel,'' \emph{submitted to IEEE Transactions on Information Theory}, pp.
  12 -- 17, 2013.

\bibitem{Cadambe}
V.~R. Cadambe and S.~A. Jafar, ``Interference alignment and degrees of freedom
  of the {K}-user interference channel,'' \emph{IEEE Transactions on
  Information Theory}, vol.~54, no.~8, pp. 3425--2441, Aug. 2008.

\bibitem{Maddah}
M.~Maddah-Ali, A.~Motahari, and A.~Khandani, ``Communication over {MIMO} {X}
  channels: Interference alignment, decomposition, and performance analysis,''
  \emph{IEEE Transactions on Information Theory}, vol.~54, no.~8, pp.
  3457--3470, Aug. 2008.

\bibitem{Jafar}
S.~A. Jafar and S.~Shamai, ``Degrees of freedom region of the {MIMO} {X}
  channel,'' \emph{IEEE Transactions on Information Theory}, vol.~54, no.~1,
  pp. 151--170, Jan. 2008.

\bibitem{Wang3}
R.~Wang and X.~Yuan, ``{MIMO} multiway relaying with pairwise data exchange: A
  degrees of freedom perspective,'' \emph{http://arxiv.org/abs/1401.7229v1},
  Submitted on 28 Jan 2014.

\bibitem{Cover}
T.~Cover and A.~Gamal, ``Capacity theorems for the relay channel,'' \emph{IEEE
  Transactions on Information Theory}, vol.~25, no.~5, pp. 572--584, Sep. 1979.

\bibitem{Chaaban}
A.~Chaaban, A.~Sezgin, and A.~S. Avestimehr, ``Approximate sum-capacity of the
  y-channel,'' \emph{IEEE Transactions on Information Theory}, vol.~59, no.~9,
  pp. 5723--5740, Sep. 2013.

\bibitem{Jafar1}
S.~A. Jafar, \emph{Interference Alignment - A New Look at Signal Dimensions in
  a Communication Network}.\hskip 1em plus 0.5em minus 0.4em\relax Foundations
  and Trends in Communications and Information Theory, 1-136, vol.~7, no.~1.

\end{thebibliography}

\end{document}